\documentclass[english,aps,pra,twocolumn,floats,noshowpacs,floatfix]{revtex4}
\pdfoutput=1
\usepackage[T1]{fontenc}
\usepackage[latin1]{inputenc}
\usepackage{graphicx}
\usepackage{amssymb}

\makeatletter

\usepackage{ifthen}

\usepackage{ifpdf}

\usepackage{color}

\usepackage{bm}

\usepackage{latexsym}

\usepackage{hyperref}

\ifpdf

\usepackage{epstopdf}

\else

\usepackage{epsfig}

\fi

\newcommand{\be}[1]{\begin{eqnarray}\ifthenelse{#1=-1}{\nonumber}{\ifthenelse{#1=0}{}{\label{e#1}}}}
\newcommand{\ee}{\end{eqnarray}} 

\newcommand{\half}{\mbox{\small $\frac{1}{2}$}}
\newcommand{\abs}[1]{\left|#1\right|}

\usepackage{babel}
\makeatother

\begin{document}

\title{Avalanches of Bose-Einstein Condensates in Leaking Optical Lattices}

\author{G. S. Ng$^{1}$, H. Hennig$^{2,3}$, R. Fleischmann$^{2}$, T. Kottos$^{1,2}$,
and T. Geisel$^{2,3}$ }

\affiliation{$^{1}$Department of Physics, Wesleyan University, Middletown, Connecticut
06459, USA \\
 $^{2}$MPI for Dynamics and Self-Organization, Bunsenstraß e 10,
D-37073 Göttingen, Germany\\
 $^{3}$Institute for Nonlinear Dynamics, University of Goettingen,
37073 Goettingen, Germany}

\maketitle

\textbf{One of the most fascinating experimental achievements of the
last decade was the realization of Bose-Einstein Condensation (BEC)
of ultra-cold atoms in optical lattices (OL's) \cite{AK98,JBCGZ98,OTFYK01,GMEHB02}.
The extraordinary level of control over these structures allows us
to investigate complex solid state phenomena \cite{GMEHB02,CBFMMTSI01,LFMWFI05,SDKEASZL05,CVHRBSGSA05,DZSZL03}
and the emerging field of {}``atomtronics'' promises a new generation
of nanoscale devices. It is therefore of fundamental and technological
importance to understand their dynamical properties. Here we study
the outgoing atomic flux of BECs loaded in a one dimensional OL with
leaking edges, using a mean field description provided by the Discrete
Non-Linear Schrodinger Equation (DNLSE). We demonstrate that the atom
population inside the OL decays in avalanches of size $J$. For intermediate
values of the interatomic interaction strength their distribution
$P(J)$ follows a power law i.e. ${\cal P}(J)\sim1/J^{\alpha}$ characterizing
systems at phase transition. This scale free behaviour of ${\cal P}(J)$
reflects the complexity and the hierarchical structure of the underlying
classical mixed phase space. Our results are relevant in a variety
of contexts (whenever DNLSE is adequate), most prominently the light
emmitance from coupled non-linear optics waveguides \cite{KW03}.}

An optical lattice with a controlled leakage of the atomic BEC can
be realized experimentally by the action of two separate continous
microwave or Raman lasers to locally spin-flip BEC atoms (at the edges
of the OL) to a nontrapped state \cite{LFO06,BHE99}. The spin-flipped
atoms do not experience the magnetic trapping potential and are released
through gravity in two atomic beams at the ends of the OL. The mathematical
model (see Method section) that describes the dynamics of the BEC
in a leaking OL of size $M$ is \begin{eqnarray}
i\frac{\partial{\psi_{n}}}{\partial\tau} & = & \chi\abs{\psi_{n}}^{2}\psi_{n}-\half[\psi_{n-1}(1-\delta_{n,1})+\psi_{n+1}(1-\delta_{n,M})]\nonumber \\
 & - & i\gamma\psi_{n}[\delta_{n,1}+\delta_{n,M}];\quad n=1,\cdots,M\label{eq:mot}\end{eqnarray}
where $\gamma$ describes the atomic losses and $N_{n}=|\psi_{n}|^{2}$
is the atomic population at site $n$. Below we study the decay (due
to leakage) of the total atom population $N(\tau)=\sum_{n}N_{n}(\tau)$
inside the OL as a function of the initial effective interaction strength
i.e. $\Lambda=\chi N(\tau=0)/M$.

An exciting result appearing in the frame of nonlinear lattices is
the existence of stationary, spatially localized solutions, termed
Discrete Breathers (DB), which emerge due to the nonlinearity and
discreteness of the system. DBs were observed in various experimental
setups \cite{SBLSSBWS99,SES99,TMO00,SHSICC03,ESMBA98,FSEC03} while
their existence and stability were studied thoroughly during the last
decade \cite{FW98,BKR03,KRB01,CFK04,ST88}. Their importance was
already recognized in \cite{TA96} where it was shown that they act
as virtual bottlenecks which slow down the relaxation processes in
generic nonlinear lattices \cite{TA96,ST88,PLL01}. Further works
\cite{PLL01} established the fact that absorbing boundaries take
generic initial conditions towards self-trapped DB's. Recently the
same scenario was proposed for a BEC in a leaking OL where it was
observed \cite{LFO06} that $N(\tau)$ decays in sudden bursts $J$
(see inset of Fig.~1). Here, for the first time we present a full
theoretical study of the decay process of $N(\tau)$ and analyze the
distribution ${\cal P}(J)$. We find three types of dynamical behaviour
\cite{HNFKG08}: For very weak interactions $\Lambda$, the population
decay is a smooth process which does not involve any avalanches. As
interactions are increased, avalanches are created. For strong interaction
strength, their distribution $P(J)$ is exponential. In contrast,
for intermediate values of the interaction strength, $P(J)$ follows
a power law indicating the existence of a phase transition. Below,
we will focus our analysis in this critical regime.

Since we are interested in the effects of DBs on the relaxation process,
we introduce a localization parameter ${\cal PR}$ which provides
a rough estimate of the relative number of sites that are occupied
by the remaining atoms in a leaking OL. It is defined as \begin{equation}
{\cal PR}(\tau)=\frac{[N(\tau)]^{2}}{M\sum_{n}|\psi_{n}(\tau)|^{4}}\label{PR}\end{equation}
which in case of $\gamma=0$ is the standard participation ratio.
Accordingly, the more evenly the atoms spread over the lattice, the
closer ${\cal PR}$ is to a constant of order $1$. The ${\cal PR}$
approaches two limiting values \cite{HNFKG08}: (a) ${\cal PR}=1/2$
corresponding to a random superposition of uncoupled sine waves for
$U=\Lambda=0$ (linear regime) and (b) ${\cal PR}=5/9$ corresponding
to a configuration of atoms trapped in $M$ uncoupled ($T\rightarrow0$
) wells for $\Lambda\rightarrow\infty$, which might be viewed as
the formation of ${\cal O}(M)$ DB's (multi-breather regime). In the
closed system, the transition between these limiting cases is smooth
(see lower left inset of Fig.~1).

In the open system ($\gamma>0$) ${\cal PR}$ is a time dependent
quantity. We numerically study the value ${\cal PR}$$_{S}$=${\cal PR}$$(\tau^{*})$
for large times $\tau^{*}$ when the system has evolved from its initial
thermalized state into a quasi-steady state %
\footnote{Even though these states might not be true steady states, they are
extremely long lived. For small $\Lambda<\Lambda_{b}$the quasi-steady
state looses norm but keeps its {}``shape'' leading to a constant
${\cal PR}(\tau)$. For large ratios $\Lambda\gtrsim2$ and large
system sizes the numerical solutions have not yet reached the final,
quasi-steady state and the true shape of the curves might differ in
this regime. In this letter, however, we focus solely on parameters
$\Lambda\lesssim1$. %
}. For nonzero $\gamma$, instead of a smooth transition between the
two extremes we observe a sharp drop of the ${\cal PR}_{S}$ at a
critical interaction strength of $\Lambda_{b}\approx0.15$ resembling
a phase transition (our numerics indicate that this transition becomes
a step function in the limit $M\rightarrow\infty$). ${\cal PR}_{S}$
drops down to its lowest possible value $(\propto1/M$) corresponding
to a single occupied site, i.e. the final state consists of a one
self-trapped DB. If $\Lambda$ increases further (Figs. 2b-c) the
final state consists of a successively larger number of breathers,
while for some critical $\Lambda$ it evolves into a more complicated
state with a power-law distribution of norms $N_{n}$ (see Fig.~1
left upper inset) %
\footnote{These scale invariant states might be transient and evolve into multi
breather states; they are, however, at least exponentially (with system
size) long lived.%
}.

In the following it is important to realize that if a breather solution
exists for some value of $\Lambda$, it exists for all $\Lambda's>0$
(and for large enough $M$). This conclusion can be easily drawn by
noting that a self-trapped DB is not directly coupled to the leaking
edges, thus we can assume $\gamma=0$ and then appropriately scale
Eq.~\ref{eq:mot}. Therefore breather solutions in particular do
exist for $\Lambda<\Lambda_{b}$ as well. So, what is the nature of
the sharp transition at $\Lambda_{b}$? To answer this question, we
have to examine the thermalized random initial states more closely.
For small $\Lambda\approx0$ these states have exponentially distributed
norms $P_{N_{n}}(x)=M$$exp(-Mx)$ \cite{HNFKG08}. In addition,
our numerical calculations (see the phase space analysis below) indicate
that the minimal total nonlinearity associated with the central site
of a self-trapped DB solution satisfy the relation $M\Lambda\left|\psi\right|^{2}\thickapprox1$
. Thus, a necessary condition in order to excite such a DB is that
at least one site of the thermalized initial state has a norm $x>x_{0}=1/(M\Lambda)$.
Assuming independent random norms for the individual sites we get
that the probability for such an event is $W_{M}(x)=1-(1-exp(-Mx))^{M}$.
By increasing $M$, we find that $W_{M}(x)$ posseses a steep transition
from $1$ to $0$ at approximately $x_{1/2}$ with $W_{M}(x_{1/2})=1/2.$
We can thus give a rough estimation of the lower bound for $\Lambda_{b}$
by demanding $x_{1/2}\geq x_{0}$, which leads to

\begin{equation}
\Lambda_{b}\geq-\frac{1}{\ln\left(1-\left[0.5\right]^{1/M}\right)}\approx\frac{1}{\ln M-\ln\ln2}\label{lambdab}\end{equation}
in fair agreement with the value $\Lambda_{b}\approx0.15$ for the
system sizes we studied %
\footnote{Note that this is only a necessary condition and the fact that this
lower bound vanishes logarithmically in the thermodynamic limit does
not imply that $\Lambda_{b}$vanishes as well. Indeed our numerics
indicate a convergence to a value near the measured one.%
}.

The most interesting situation emerges for large enough system sizes
and (for the system sizes we could numerically study) interatomic
interactions in the range of $\Lambda\approx0.5-1$ %
\footnote{We speculate that this regime grows down to $\Lambda_{b}$with increasing
system sizes%
}. In this case we have found that the atoms leak out of the lattice
in avalanches following a scale-free distribution, i.e. ${\cal P}(J)\sim J^{-\alpha}$,
while the norms $N_{n}$ at individual lattice sites are distributed
as $P_{N_{n}}(x)\sim x^{-\beta}$with $\alpha\approx\beta$ (see Fig.
1 left upper inset). This power law behavior for a whole range of
parameter values $\Lambda$ could be interpreted as a signature of
a self-organized critical state \cite{J98}.

Let us study in more detail the dynamics that lead to the creation
of an avalanche in this parameter regime. One such event is depicted
in Fig.~2d. A moving DB, coming from the bulk of the lattice, collides
with the outer most self-trapped DB. As a result the self-trapped
breather is shifted inwards by one site while at the same time a particle
density proportional to the density of the moving breather tunnels
through the self-trapped DB. Eventually this particle density will
reach the leaking edge of the OL and decay in a form of an avalache. 

We have also found numerically \cite{HNFKG08} that during the migration
process of the self-trapped DB, the number of particles and the energy
of the three lattice sites involved is conserved, thus allowing us
to turn the problem to the analysis of a reduced $M=3$ system with
interaction strengths $\Lambda$ in the critical range. As shown in
Fig. 3a (inset), the non-linear trimer exhibits a hierarchical mixed
phase space structure with islands of regular motion (tori) embedded
in a sea of chaotic trajectories. Trajectories inside the islands
correspond to self-trapped DBs, provided that their frequency is outside
the linear spectrum. In contrast, chaotic trajectories have continuous
Fourier spectra, parts of which overlap with the linear spectrum of
the infinite lattice \cite{FW98}. As long as the self-trapped DB
is stable, it acts as a barrier which prevents atoms from reaching
the leaking boundary. Thus, a necessary condition for an avalanche
event is the destabilization of the DB. This is caused by a lattice
excitation with particle density $N_{1}^{pert}$greater than a critical
value which can push the regular orbit out of the island. At the same
time a portion $N_{3}^{max}\propto N_{1}^{pert}$transmits through
(see Fig. 3a). As can be seen from Fig. 2, the migration process usually
moves the DB inwards by one site which correspond to another island
in the phase space of the extended lattice. Details of the migration
process are still under investigation \cite{HNFKG08,R04}.

We conjecture that the size of avalanches $J$ is proportional to
the size $S$ of islands in the mixed phase space found in the Poincaré
section of the non-linear trimer. A heuristic model \cite{HNFKG08}
that mimics the hierarchical (`island-over island') structure of a
typical mixed phase space leads us to a power law distribution $P\mathrm{(S)\sim S^{-\alpha}}$
where $1<\alpha<3$. The lower bound comes from the requirement of
having an infinite number of islands (self-similar property) while
the upper-bound results from the requirement that the total volume
of the phase space be finite. To verify the above prediction for $P(S)$
is a computationally demanding task. Therefore we used a numerically
more convenient model: the kicked rotor which is a paradigmatic model
of mixed phase space dynamics \cite{LR04}. Leaving aside the technical
details \cite{HNFKG08}, we present in Fig. 3b the outcome of our
numerical calculations. The results confirm the power law scaling
of the island sizes over more than three orders of magnitude thus
confirming the validity of our conjecture. 

In conclusion, we have shown that there is a critical regime of interatomic
interactions, where particles are ejected out of a leaky OL in scale
free avalanches. The observed power law distribution, is dictated
by the hierarchical structure of the mixed phase space. Our results
are quite universal and should be observable in many other experimental
realizations of the DNLSE including molecular crystals, globular proteins
and non-linear optics.

\subsection*{METHODS}

The simplest model that captures the dynamics of a dilute gas of bosonic
atoms in a deep OL with chemical potential small compared to the vibrational
level spacing, is the Bose-Hubbard Hamiltonian. In the case of weak
interatomic interactions (superfluid limit) and/or a large number
of atoms per well (so that the total number of atoms $N\sim{\cal O}(10^{4}-10^{5})$
is much bigger than the number of wells $M$), a further simplification
is available since the BECs dynamics admits a semiclassical mean field
description \cite{TS01}. The resulting Hamiltonian is \begin{equation}
{\cal H}=\sum_{n=1}^{M}[U|\psi_{n}|^{4}+\mu_{n}|\psi_{n}|^{2}]-\frac{T}{2}\sum_{n=1}^{M-1}(\psi_{n}^{*}\psi_{n+1}+c.c.)\label{eq:H0_livi}\end{equation}
 where $n$ is the well index, $|\psi_{n}(t)|^{2}\equiv N_{n}(t)$
is the mean number of bosons at well $n$, $U=4\pi\hbar^{2}a_{s}V_{{\rm eff}}/m$
describes the interaction between two atoms on a single site ($V_{{\rm eff}}$
is the effective mode volume of each site, $m$ is the atomic mass,
and $a_{s}$ is the $s$-wave atomic scattering length), $\mu_{n}$
is the well chemical potential, and $T$ is the tunneling amplitude.
The \char`\"{}wavefunction amplitudes\char`\"{} $\psi_{n}(t)\equiv\sqrt{N_{n}(t)}\exp(-i\phi_{n}(t))$
can be used as conjugate variables with respect to the Hamiltonian
$i{\cal H}$ leading to a set of canonical equations $i\partial_{t}\psi_{n}=\partial{\cal H}/\partial\psi_{n}^{*};i\partial_{t}\psi_{n}^{*}=-\partial{\cal H}/\partial\psi_{n}$.
Substituting (\ref{eq:H0_livi}) we get the DNLSE. To simulate the
leaking process at the two edges, we supplement the standard DNLSE
with a local dissipation at the two edges of the lattice. The resulting
leaking DNLSE is \cite{LFO06} \begin{eqnarray}
i\frac{\partial{\psi_{n}}}{\partial\tau} & = & (\chi\abs{\psi_{n}}^{2}+\tilde{\mu}_{n})\psi_{n}-\half[\psi_{n-1}(1-\delta_{n,1})+\psi_{n+1}(1-\delta_{n,M})]\nonumber \\
 & - & i\gamma\psi_{n}[\delta_{n,1}+\delta_{n,M}];\quad n=1,\cdots,M\label{eq:mot}\end{eqnarray}
where the normalized time is defined as $\tau=Tt$, $\chi=2U/T$ is
the rescaled nonlinearity, $\tilde{\mu}_{n}=\mu_{n}/T$ is the rescaled
chemical potential and $\gamma$ is the atom emission probability
describing atomic losses. It is useful to define the initial effective
interaction per well i.e. $\Lambda=\chi\rho$ where $\rho=N(\tau=0)/M$
is the initial average density of atoms in the OL. In our numerical
experiments we have used initial conditions with randomly distributed
phases, and an almost constant amplitude with only small random fluctuations
across the OL. We normalized the wave functions such that $N(\tau=0)=1$.
The initial states are first thermalized during a conservative (i.e.
$\gamma=0$) transient period of, typically, $\tau=500$. The dissipation
at the lattice boundaries is switched on only after this transient
is completed, leading to a progressive loss of atoms. We study the
decay and the statistical properties of $N(\tau)$ as a function of
the parameter $\Lambda$. The results reported in this Letter correspond
to $\tilde{\mu}_{n}=0$ and dissipation rate $\gamma=0.2$ which is
within the experimentally accessible range \cite{LFO06}. Nevertheless
we have checked that we get the same qualitative behavior for other
values of $\gamma$ \cite{HNFKG08}.


\ \\
 \ \\

\begin{figure}[h]
\includegraphics[width=1\columnwidth]{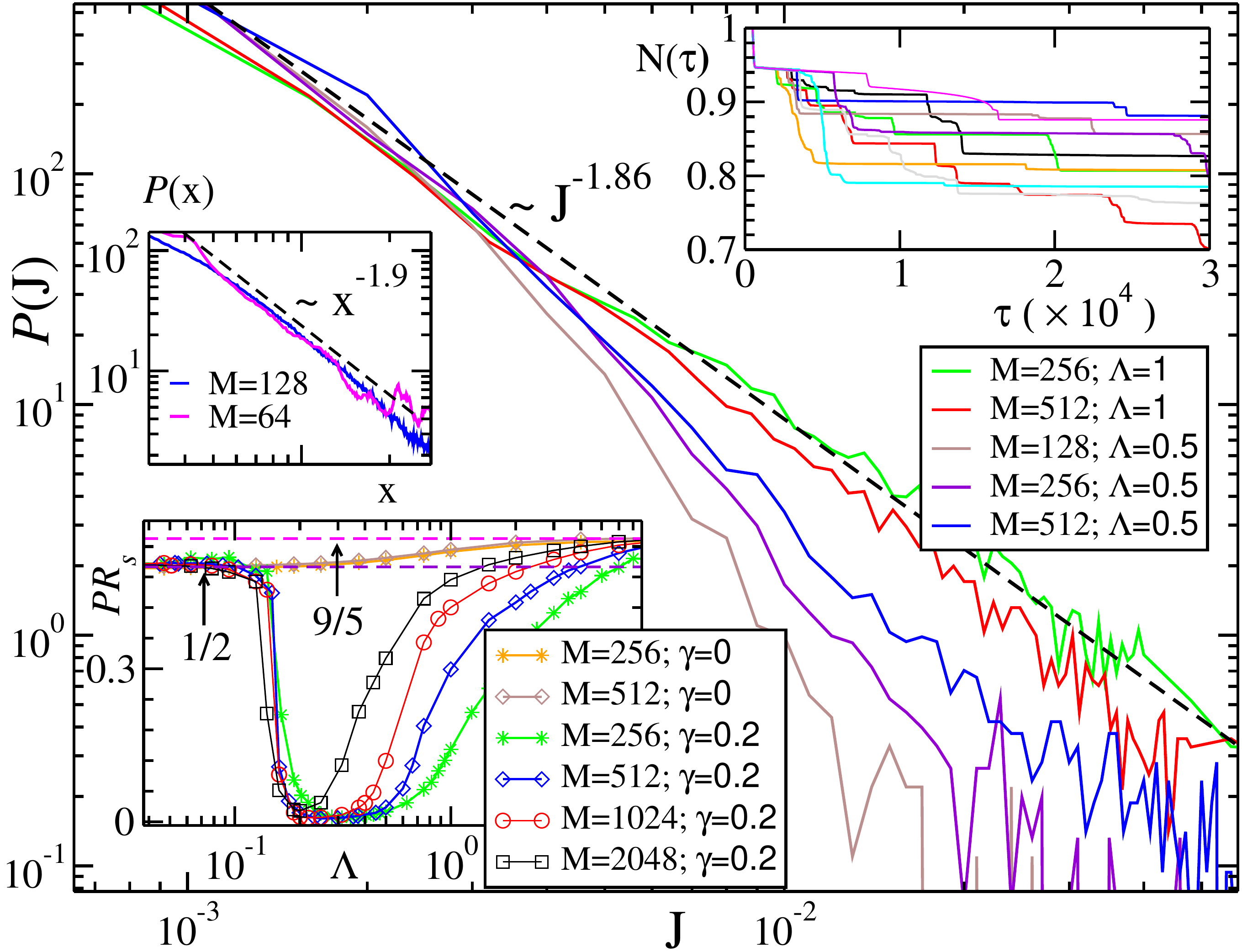} 

\caption{Main panel: Distribution of avalanches $\mathcal{\mathcal{P}\mathrm{(J)}}$
of various system sizes $M$ for interatomic interaction strengths
$\Lambda=0.5$ and $\Lambda=1$. In the former case we observe a convergance
to a power law distribution $\mathcal{\mathcal{P}\mathrm{(J)}}\sim J^{-\alpha}$
as the lattice size $M$ increases, while in the latter case the asymptotic
distribution has already been reached for $M=512.$ The best least
square fit indicates that $\alpha=1.86\pm0.04$ in agreement with
the bounds $1<\alpha<3$ (see discussion at the text). The distribution
is generated over different initial thermal excitations. Upper right
inset: Representative realizations of atomic population decay showing
avalanches. Upper left inset: Power law distribution of norms $\mathcal{\mathcal{P}\mathrm{(x=|\psi_{n}|^{2})\sim x^{-\beta}}}$
for $\Lambda=1$. The best least square fit indicates that $\beta=1.9\pm0.05\approx\alpha.$
Lower left inset: The localization parameter $\mathcal{PR}_{S}$ as
a function of the initial effective interatomic interaction $\Lambda.$ }

\end{figure}

\begin{figure}[h]
\includegraphics[clip,width=1\columnwidth]{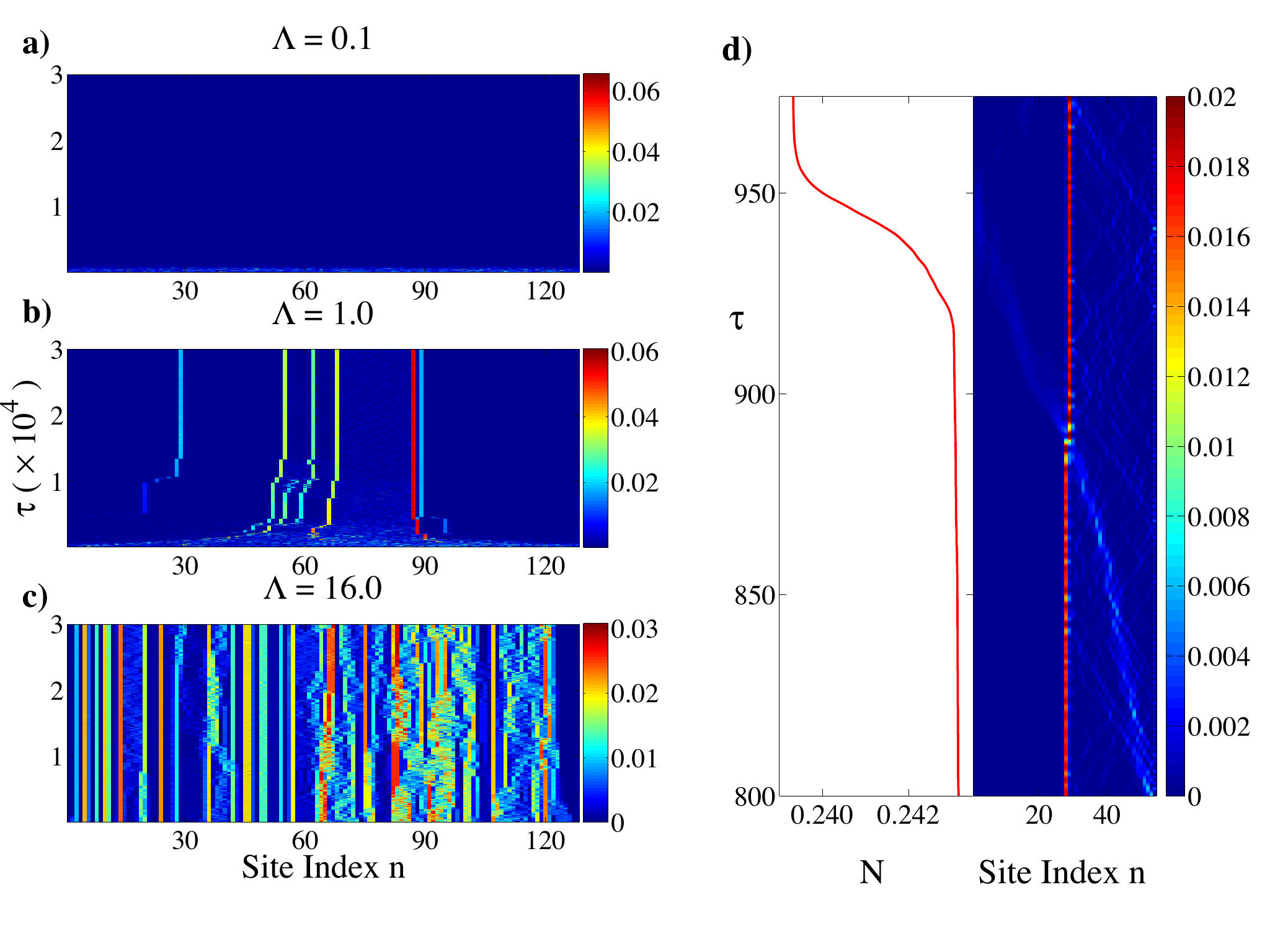} 

\caption{Left panel: Evolution of atomic population for a lattice of size $M=128$
and various interatomic interaction strengths $\Lambda:$ (a) $\Lambda=0.1<\Lambda_{b}$
where no DBs are formed; (b) $\Lambda=1>\Lambda_{b}$ corresponding
to the critical regime where scale free avalanches are created; (c)
$\Lambda=16\gg1$ corresponding to the creation of $\mathcal{\sim O}(M)$
DB's (multibreather regime); (d) Snapshot of an avalanche event. On
the left subpanel, we are plotting $\tau$ vs $N(\tau)$ whereas on
the right we are reporting a representative collision event between
the outer most self-trapped DB and a lattice excitation (moving breather)
that leads to destabilization of the DB. During the collision, part
of the moving breather tunnels through the self-trapped DB and travels
towards the edge of the lattice. The arrival of the transmitted density
at the edge registers an avalanche in the atomic population $N(\tau)$
(see left subpanel). }

\end{figure}
\begin{figure}[h]
\includegraphics[width=1\columnwidth]{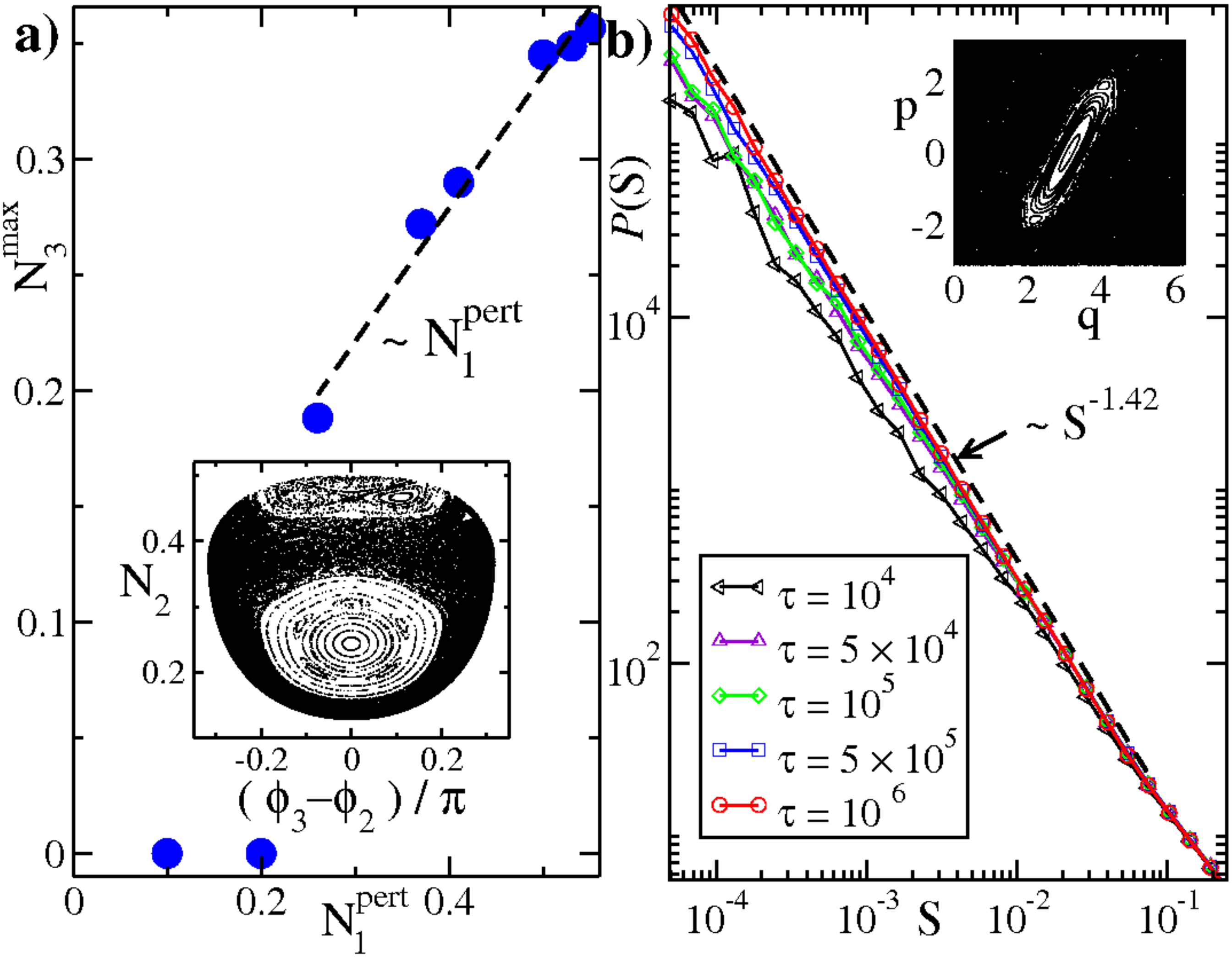} 

\caption{\textbf{(a)} A destabilization process of a DB hosted by a three site
system. In the inset we report a Poincare section of the phase space
by referring to the variables $(N_{2},\psi)$, where $N_{2}$ is the
norm at site 2 (where the main part of the DB is located), and $\psi=\phi_{3}-\phi_{2}$
is the phase missmatch between the center, and the next site. In the
main figure, we report the outgoing atomic population measured at
site three $N_{3}^{max}$(associated with an avalanche event--see
Fig. 2) versus the incoming atomic population $N_{1}^{pert}$ from
the first site. We observe that atomic population tunnels through
the DB only if $N_{1}^{pert}\geq0.25$, corresponding to the minimal
excitation needed to trigger the destabilization of the self-trapped
DB. \textbf{(b)} The distribution ${\cal P}(S)$ of island-sizes for
the kicked rotor with kicking strength $K=3.5$, corresponding to
a mixed phase space (inset). The island sizes were evaluated by studing
the evolution of close-by initial conditions as a function of their
seperation $S$ for succesively longer times $\tau$ . We see that
in the limit of $\tau\rightarrow\infty$ the distribution converge
to a power law.}

\end{figure}

\end{document}